\documentclass[conference]{IEEEtran}
\IEEEoverridecommandlockouts
\usepackage{cite}
\usepackage{amsmath,amssymb,amsfonts}
\usepackage{algorithmic}
\usepackage{graphicx}
\usepackage{textcomp}
\usepackage{xcolor}
\usepackage[ruled,linesnumbered]{algorithm2e}
\usepackage{subfigure}
\usepackage{booktabs}
\usepackage{multirow}
\usepackage{array}
\def\BibTeX{{\rm B\kern-.05em{\sc i\kern-.025em b}\kern-.08em
    T\kern-.1667em\lower.7ex\hbox{E}\kern-.125emX}}
\begin{document}

\title{Improving Speech Representation Learning via Speech-level and Phoneme-level Masking Approach}

% \and
% \and
% \and
% \and
% \and

\author{\IEEEauthorblockN{Xulong Zhang$^1$, Jianzong Wang$^{1\ast}$\thanks{$^\ast$Corresponding author: Jianzong Wang, jzwang@188.com.}, Ning Cheng$^1$, Kexin Zhu$^{1,2}$, Jing Xiao$^1$}
\IEEEauthorblockA{$^1$\textit{Ping An Technology (Shenzhen) Co., Ltd.}}
\IEEEauthorblockA{$^2$\textit{School of Software, Fudan University, Shanghai, China}}
}

\maketitle

\begin{abstract}
    Recovering the masked speech frames is widely applied in speech representation learning. However, most of these models use random masking in the pre-training. In this work, we proposed two kinds of masking approaches: (1) speech-level masking, making the model to mask more speech segments than silence segments, (2) phoneme-level masking, forcing the model to mask the whole frames of the phoneme, instead of phoneme pieces. We pre-trained the model via these two approaches, and evaluated on two downstream tasks, phoneme classification and speaker recognition. The experiments demonstrated that the proposed masking approaches are beneficial to improve the performance of speech representation.
\end{abstract}


\section{Introduction}

%[Self-supervised learning in ASR]
Speech representation learning has proved to be an effective method of extracting high-level speech information~\cite{chung2016audio}. The speech representation learning is usually based on pre-training on a large scale unlabeled speech data. It is well known that labeled speech data requires huge amounts of labour costs, while the unlabeled corpus is relatively easy to obtain~\cite{zhang2022TDASS}. The extracted speech representation could be used to improve the downstream speech and language processing (SLP) tasks.

Several Self-supervised learning (SSL) methods have been proposed for speech representation learning~\cite{2021Wav2vec,tang2022avqvc}. Autoregressive predictive coding (APC)~\cite{chung2019unsupervised} is designed to learn an auto-regressive model by predicting future speech frames. Contrastive predictive coding (CPC)~\cite{Oord2018Representation} uses a contrastive loss that maximizing the mutual information between present representations and future signals. Wav2vec~\cite{Schneider2019wav2vec} also learns a contrastive objective that distinguishs the true future audio from negative samples. All of the above SSL models could extract speech representation through pre-training on large amounts of unlabeled data.

%[Current problems: 1. silence 2. the first or last token that is redundant]
Masked language model (MLM)~\cite{Wang2020Unsupervised} is another popular SSL architecture of speech representation learning. The MLM model often uses a Transformer-based network to reconstruct the masked or altered speech frames. Masked predictive coding (MPC)~\cite{jiang2019improving} used a Transformer encoder to predict the masked filter bank (Fbank) features, and fine-tuned the Transformer decoder for transcript prediction. Mockingjay~\cite{Liu2020Mocking} proposed to use a BERT-style masking strategy with random selected frames for pre-training. Transformer encoder representations from alteration (TERA)~\cite{liu2020tera} extended the work of Mockingjay, introducing three auxiliary multi-task learning objectives (temporal, channel, and magnitude) to the self-supervised learning for speech. Wav2vec2.0~\cite{Baevski2020wav2vec2} masks the speech inputs in the latent space, and pre-trains the model by a contrastive loss of the quantized latent speech representation. 

%[Motivation and our method]
%first step: eliminate silence; second step: eliminate redundant tokens (easy task)
%[Main contributions]
Despite the impressive performance of these MLM models, some pre-training strategies still could be improved. One is that most of the masking approaches is random masking for these models. They do not consider any prior knowledge in the speech. In natural language processing (NLP), some previous works were proposed to use knowledge based masking strategies, instead of random masking. Enhanced Representation through kNowledge IntEgration (ERNIE)~\cite{Zhang2019ERNIE} is designed to learn language representation by entity-level masking and phrase-level masking. SpanBERT~\cite{Joshi2020SpanBERT} proposed to mask contiguous spans and designed a span boundary objective loss relying on the relative position within the masked span. Cui et al.~\cite{Cui2019Pre} presented whole word masking to BERT~\cite{Devlin2018BERT}, providing a more challenging pre-training task of predicting all the characters in a complete Chinese word. RoBERTa~\cite{Liu2019RoBERTa} found that dynamic masking is beneficial to pre-training on large datasets, which generates the masking pattern dynamically when feeding it to the model.

For SLP tasks~\cite{Jia2020Large}, random masking approach is likely to select the non-speech segments. Some speech data may contain lots of silence like telephone conversational corpus. These less informative segments make the pre-training task too easy to recover these non-speech frames. Inspired by the above works in NLP, we proposed two levels of masking approaches to the speech representation learning: (1) speech-level masking, (2) phoneme-level masking. The speech-level approach will mask more speech segments than silence segments at the pre-training stage. We think the speech frames may contain more useful acoustic information than non-speech frames. The voice activity detection (VAD) algorithm~\cite{Salishev2016Voice} is applied to split the original speech into speech segments or not. Furthermore, the phoneme-level approach will force the model to reconstruct the whole frames of the masked phoneme. This is a more challenging task than just reconstructing some pieces of the phoneme. A pre-trained automatic speech recognition (ASR) model~\cite{Kim2017Joint} is used to force align each speech frame to a corresponding phoneme. We also combined the speech-level and phoneme-level masking together, to obtain better speech representation in the pre-training and better performance in downstream tasks.

\section{Methodology}

\subsection{Model Architecture}

This paper exploits the transformer-based masked language model (MLM) as the overall model architecture. Our works mainly focus on the masking approach in temporal channel at the pre-training stage. A masking approach alters or masks a number of speech frames of the original input $X=(x_1,x_2,...,x_T)$. The masking sequence is $M=(m_1,m_2,...,m_T)$, where $T$ is the length of acoustic sequence. We denote the masking process as conducting element-wise product $\odot$ between $X$ and $M$. Then, the MLM $P_{MLM}$ predicts an output $\widetilde{X}=(\widetilde{x}_1,\widetilde{x}_2,...,\widetilde{x}_T)$ based on the masked input.

The objective of pre-training is to minimize the error between the predicted output $\widetilde{X}$ and the original input $X$. As in Mockingjay~\cite{Liu2020Mocking} and TERA~\cite{liu2020tera}, we also used the $\mathcal{L}_{1}$ loss as follows:
\begin{equation}
	\mathcal{L}_{1} = | X - \widetilde{X}| 
\end{equation}

%One common masking approach in temporal channel is random masking, which selects the masked tokens randomly in time domain. 

%Suppose a series of acoustic features are as input $X=(x_1,x_2,...,x_T)$, where T is the length of time-domain signal. A masking series with equal time length $M=(m_1,m_2,...,m_T)$ is used to mask certain tokens in $X$ by conducting element-wise product with $X$. The CMLM $P_{CMLM}$ takes the product result as input and generates a predicted output $\widetilde{X}$:
%
%\begin{equation}
%\widetilde{X}= P_{CMLM}(X \odot M)
%\end{equation}

%where $$m_{1:T}=\left\{
%\begin{aligned}
%0 & ,  \text{if the corresponding tokens are randomly masked,} \\
%1 & ,  \text{otherwise.}
%\end{aligned}
%\right.$$

%The CMLM tries to minimize the distance between $X$ and $\widetilde{X}$ with respect to L1 loss:
%
%\begin{equation}
%\mathcal{L}_{1} = | X - \widetilde{X}| 
%\end{equation}

The most common masking approach in temporal channel is random masking, which selects the masked frames randomly in time domain. As in \cite{Liu2020Mocking}, the approach masks successive $C$ frames once they randomly pick a frame as the starting point. This is to avoid the model utilizing local smoothness of the acoustic frames. %The random masking approach is described in Algorithm~\ref{random}, generating the random masking sequence $M_R$.

% \LinesNumbered
% \begin{algorithm}[!htbp]
% 	\SetAlgoLined
% 	\KwIn{Acoustic features $X=(x_1,x_2,...,x_T)$, Pre-trained ASR model $P_{ASR}$}
% 	\KwOut{Phoneme masking $M_P=(m_1,m_2,...,m_T)$}
% 	\BlankLine
	
% 	\For{$1 \leq i\leq T$}{
% 		$m_i = 1$
% 	}
% 	\textit{Predict} the text $Y=(y_1,y_2,...,y_L)$ from $X$: $Y = P_{ASR}(X)$\\
% 	\textit{Force alignment} $X$ and $Y$ to get aligned phonemes $Y^{\prime}=(y^{\prime}_1,y^{\prime}_2,...,y^{\prime}_T)$\\
% 	\For{$1 \leq i \leq L$}{
% 		\textit{Record} begin index of $y_i$ in $Y^{\prime}$ to a list $B$\\
% 		\textit{Record} end index of $y_i$ in $Y^{\prime}$ to a list $E$\\
% 	}
% 	%Randomly select $N$ ($\leq$ T) tokens to mask;\\
% 	\For{$1 \leq j \leq N$}{
% 		\textit{Sample} a random phoneme $y^{\prime\prime}_j \in [y_1,y_S]$ \\
% 		\textit{Select} the begin index $b_j$ of $y^{\prime\prime}_j$ from list $B$ \\
% 		\textit{Select} the end index $e_j$ of $y^{\prime\prime}_j$ from list $E$ \\
% 		$m_{b_j}=m_{{b_j}+1}=...=m_{e_j}=0$
% 	}
% 	%	$M_P=M$;\\
% 	\Return $M_P$
% 	\caption{Phoneme-Level Masking Approach}
% 	\label{phoneme}
% \end{algorithm}

\subsection{Speech-Level Masking Approach}

We firstly proposed a speech-level masking approach based on the VAD algorithm~\cite{Salishev2016Voice}. The VAD algorithm is a binary classification problem determining whether an input signal contains speech or not. Certain features (e.g. energy, cepstral coefficient, etc.) are extracted from a segment of the input audio signals. Then, a threshold $\theta$ is set to classify the segment as the speech if the value of extracted features exceeds the threshold $\theta$.

%As discussed in last section, masking non-speech segment can hardly help our model increase its generalization ability. We decide to mask speech segment most of the time.
We think masking more speech segments could help the model to learn more useful acoustic information. A ratio parameter $\rho$ is set to control the proportion of masked speech and non-speech segments. The reason why a small portion of non-speech segments are still masked is that the silence may sometimes contain high semantic knowledge. We firstly used VAD algorithm to classify whether each frame is speech or not. Then, the starting points of masking are selected from speech list $A$ or non-speech list $B$, according to the ratio $\rho$. After that, $C$ successive frames after each starting point are also masked, to generate the speech-level masking sequence $M_S$. %The speech-level masking approach is described in Algorithm \ref{speech}.

%If a token in speech segments is selected, we mask $C$ successive tokens, again; if a token in non-speech segments is selected, we only mask this token. Suppose the number of speech tokens and non-speech tokens are $N$ and $M$ ($1\leq N, M \leq T$), respectively. The number of selected masked tokens is $O$ ($1\leq O \leq T$). The speech-level masking approach is described in Algorithm \ref{speech}.

%Afterwards, we obtain predicted output  $\widetilde{X_S}$ using speech-level masking with the following operation:
%
%\begin{equation}
%\widetilde{X_S}= P_{CMLM}(X \odot M_S)
%\end{equation}

%The predicted output $\widetilde{X_R}$ using random masking is then as follows:
%
%\begin{equation}
%\widetilde{X_R}= P_{CMLM}(X \odot M_R)
%\end{equation}

\begin{figure}[ht]
	\centering
	\subfigure[Random Masking Speech and Silence Segments] { \label{fig1a}
		\includegraphics[width=0.9\columnwidth]{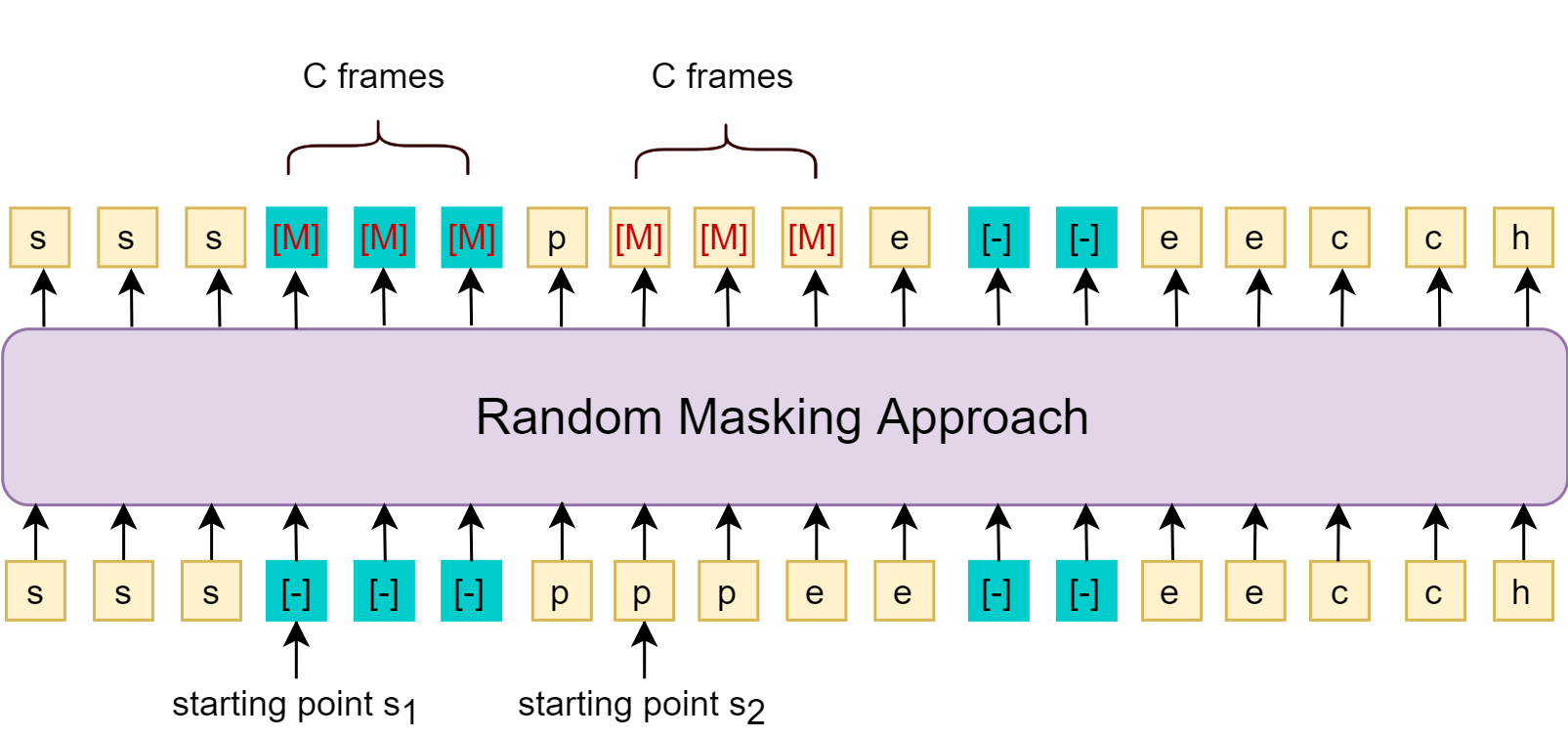}
	}
	\subfigure[Speech-Level Masking More Speech Segments] { \label{fig1b}
		\includegraphics[width=0.9\columnwidth]{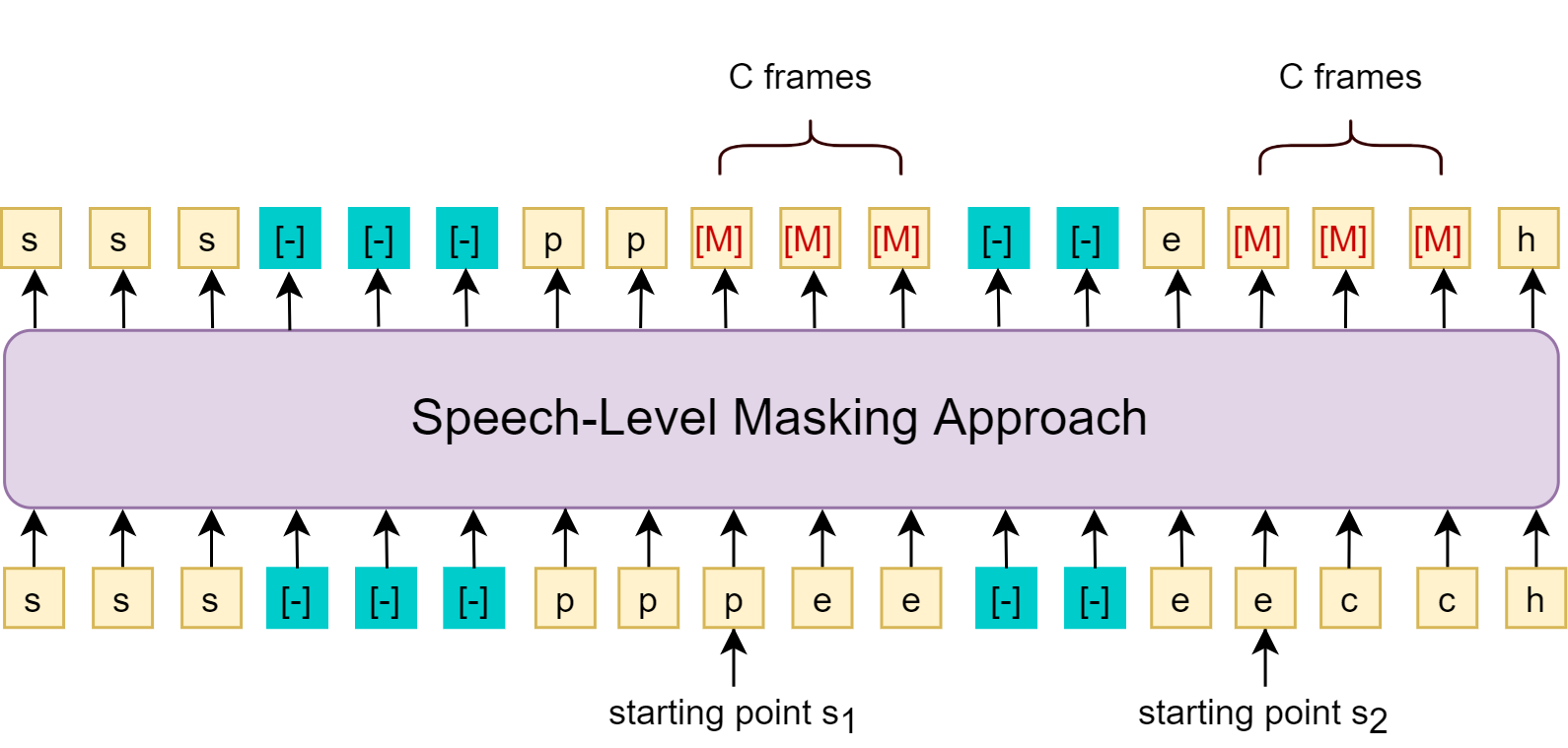}
	}
	\subfigure[Phoneme-Level Masking All Frames of Phoneme \textbf{p} and \textbf{e}] { \label{fig1c}
		\includegraphics[width=0.9\columnwidth]{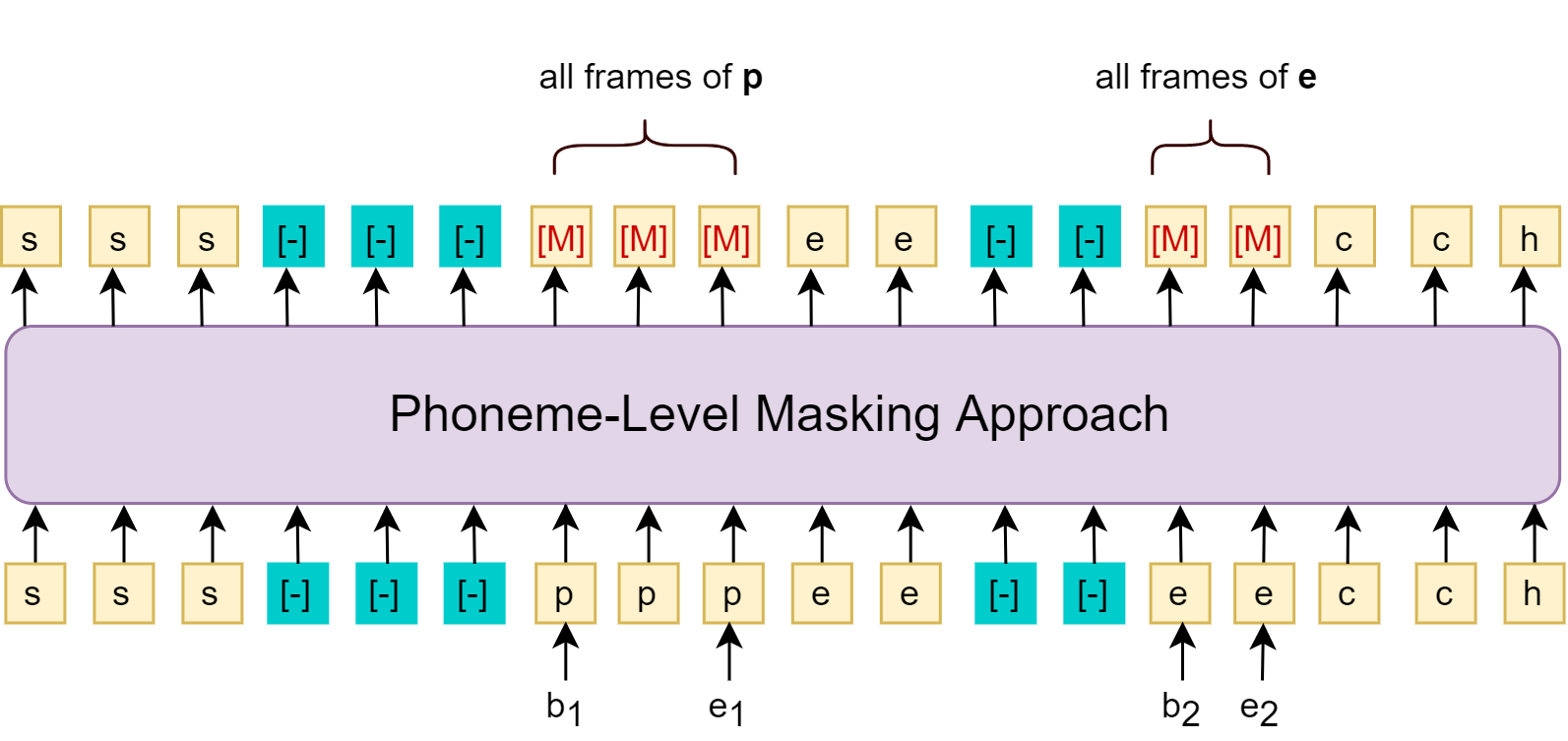}
	}
	\caption{Visualization of Different Masking Approaches}
	\label{fig1}
\end{figure}

\subsection{Phoneme-Level Masking Approach}

In this section, we proposed a more challenging task by masking the whole frames of phoneme. %The phoneme-level masking approach is described in Algorithm~\ref{phoneme}.
We firstly used a pre-trained ASR model~\cite{Kim2017Joint} $P_{ASR}$ to predict the text $Y$ of acoustic features $X$. In real application, the $P_{ASR}$ is usually trained by a small amount of labeled data. Then, we applied force alignment to map each speech frame to one phoneme, generating the aligned phoneme sequence $Y^{\prime}$. Force alignment~\cite{McAuliffe2017Montreal} is a task of determining the time boundaries between phonemes of acoustic features. After that, $N$ phonemes are selected randomly. All the frames between the begin index $b_j$ and end index $e_j$ of each selected phoneme $y^{\prime\prime}_j$ are masked. Especially, we masked the whole frames of each phoneme in this approach, instead of $C$ successive frames.

\subsection{Visualization}

We illustrated the masking process of above three masking approaches in Figure~\ref{fig1}. The input signal is the speech frames of a word $speech$ (in light yellow boxes). The non-speech frames are denoted as a symbol $[-]$ (in light green boxes). The masked frames after masking approach are denoted as a red symbol $[M]$.

\begin{table*}
	\small
	\centering
	\caption{\textit{Compared with Different Masking Approaches, Results on Librispeech, Accuracy(\%)}} 
	\vspace{1.0em}
	\setlength\tabcolsep{3pt} 
	\scalebox{0.95}{
		\begin{tabular}{|c|c|cccc|cccc|}
			
			\hline
			
			\multirow{2}{*}[-0.5ex]{Pre-training} &\multirow{2}{*}[-0.5ex]{Masking Approach} &\multicolumn{4}{c|}{\textit{train-clean-100}} & \multicolumn{4}{c|}{\textit{train-clean-360}}\\  
			
			&& {Phoneme-L} & {Phoneme-1H} & {Speaker-F} & {Speaker-U} & {Phoneme-L} & {Phoneme-1H} & {Speaker-F} & {Speaker-U}\\
			
			\hline
			\hline

			\multirow{4}{*}[-0.5ex]{Mockingjay~\cite{Liu2020Mocking}} &Random &69.6 &78.8 &68.4 &96.1 &67.5 &78.2 &86.9 &97.3\\ 
			&Speech-Level  &70.2 &79.3 &97.6 &97.2 &68.0 &78.1 &97.8 &98.3\\
			&Phoneme-Level &70.2 &79.7 &97.9 &\textbf{98.5} &67.8  &78.8 &\textbf{98.1} &\textbf{98.9}\\
			&Speech\&Phoneme-Level &\textbf{70.3} &\textbf{79.9} &\textbf{98.2} &98.2 &\textbf{68.5} &\textbf{78.9} &97.2 &98.3\\ \hline
			
			\multirow{4}{*}[-0.5ex]{TERA \cite{liu2020tera}} &Random &71.3 &79.1 &98.9  &99.2 &70.8 &79.6 &99.1 &99.3 \\
			&Speech-Level &71.5 &\textbf{80.3} &99.6 &99.3 &71.3 &\textbf{80.7} &99.1 &99.2 \\  
			&Phoneme-Level &71.4 &79.5 &\textbf{99.7} &\textbf{99.7} &71.7 &80.4 &99.3 &\textbf{99.5} \\ 
			&Speech\&Phoneme-Level &\textbf{71.8} &80.1 &99.5 &99.4 &\textbf{71.8} &80.5 &\textbf{99.3} &99.4 \\
			
			\hline 
			
	\end{tabular}}
	\label{tab1}
\end{table*}

For random masking approach (in Figure~\ref{fig1a}), two starting points $s_1$ and $s_2$ are randomly selected and $C$ successive frames are masked. The random masking is likely to select the silence segment ($s_1$ in Figure~\ref{fig1a}). For speech-level masking approach (in Figure~\ref{fig1b}), speech and non-speech frames are classified by VAD algorithm firstly. Then, most of the masked frames are selected as the informative speech frames ($s_1$ and $s_2$ in Figure~\ref{fig1b}). The speech-level approach will also mask $C$ successive frames. For phoneme-level masking approach (in Figure~\ref{fig1c}), phoneme boundaries are detected by force alignment algorithm. In which, $b_1$ and $e_1$ denote the begin and end frame of phoneme $p$, and $b_2$ and $e_2$ cover the frames of phoneme $e$. All frames of each phoneme are masked for a more challenging model pre-training.

In addition, speech-level and phoneme-level masking approach could also be combined together. Firstly, we apply the VAD algorithm to distinguish the speech and non-speech frames. Secondly, all frames of a detected phoneme are masked, when we choose a starting point in speech segment with probability $\rho$. Otherwise, fixed $C$ frames from a starting point in silence segment are masked. The experimental results of this combined masking approach will be shown in Section~\ref{sec:res}.

\section{Experiments}
\label{sec:exp}
%In this work, we focus on improving the speech representation learning with well-designed masking strategies. The proposed masking approaches are evaluated on two dowmstream tasks, phoneme classification and speaker recognition. We compared the performance of different masking methods, and also explored different configurations on proportion ratio $\rho$.

\subsection{Experimental Setup}
\label{sec:setup}
We used two subsets of Librispeech~\cite{Panayotov2015Librispeech} corpus for pre-training: the \textit{train-clean-100} and the \textit{train-clean-360}. To improve the speech representation learning with well-designed masking strategies. Therefore, the accuracy of downstream tasks are compared using different masking approaches, without changing the network architecture. Two MLM models $P_{MLM}$: Mockingjay~\cite{Liu2020Mocking} and TERA~\cite{liu2020tera}, are used to extract the speech representations from the last layer of the model after pre-training.

%We utilized two different classifiers to measure the performance of phoneme classification. Firstly, linear classifier was leveraged to measure the linear separability of phonemes. We denote this classifier as \textit{linear}. Secondly, classifier with one single hidden layer was also used, as not all the information encoded is fully linear. We denote this classifier as \textit{1-hidden}.
%
%We also evaluated the performance of speaker recognition from two perspectives: frame-wise recognition and utterance-wise recognition. For frame-wise recognition, the classifier predicts the speaker for each input frame. We denote this experiment setting as \textit{frame}. For utterance-wise recognition, an representation of each utterance over time is extracted. 

For phoneme classification tasks, we utilized linear classifier (denoted as \textit{Phoneme-L}) and classifier with one single hidden layer (denoted as \textit{Phoneme-1H}). For speaker recognition tasks, we performed frame-wise (denoted as \textit{Speaker-F}) and utterance-wise (denoted as \textit{Speaker-U}) classification to predict the speaker identity.

The input acoustic features are $80$-dimensional Fbank. We set the time masking width $C$ to $7$ frames for random approach. Following the previous works in \cite{Liu2020Mocking} and \cite{liu2020tera}, $3$-layers Transformer encoder network is used. The multi-head self-attention layer can extract feature information from multiple dimensions~\cite{2017Attention}. Each layer produces an output of the same dimension. The hidden size of intermediate feed-forward layer is $3072$ with dropout rate $0.1$. %All of the models are pre-trained with batch size $32$ on $4$ NVIDIA V100 GPU cards.
In addition, the VAD algorithm is implemented by Google WebRTC framework~\cite{WebRTC}. The force-alignment results are obtained by Kaldi recipes~\cite{Kaldi}. We conducted all the pre-training and dowmstream experiments on S3PRL toolkit~\cite{S3PRL}.
%The number of pre-training and downstream steps were both 200$k$.

\subsection{Results}
\label{sec:res}

As depicted in Table~\ref{tab1}, we illustrated the accuracy results ($20k$ pre-training steps, $20k$ downstream steps) of phoneme classification and speaker recognition with different masking approaches on Librispeech dataset. For Mockingjay model, the proposed three masking approaches (\textit{Speech-Level}, \textit{Phoneme-Level}, and \textit{Speech\&Phoneme-Level}) have higher accuracy than random masking approach, in both two downstream tasks. Especially in the phoneme classification task, the combined masking approach achieves the highest accuracy on both datasets. It achieves accuracy rates of 70.3\% and 79.9\% respectively when using the \textit{train-clean-100} dataset for the \textit{Phoneme-L} and \textit{Phoneme-1H}, while its effect on the \textit{train-clean-360} dataset is 68.5\% and 78.9\%, respectively. Compared to random masking approach, our three masking approaches have achieved very significant improvements in the accuracy of the frame-wise speaker recognition task. In this task, our three approaches achieved accuracies of 97.6\%, 97.9\% and 98.2\% respectively when using the \textit{train-clean-100} dataset, compared to 68.4\% for the random approach. Our approaches also achieved 97.8\%, 98.1\% and 97.2\% accuracies on the \textit{train-clean-360} dataset, respectively, as compared to the random approach accuracy of 86.9\%. For TERA model, all of our three masking approaches could outperform random masking approach in phoneme classification task. The proposed approaches also have comparable performance with random masking approach in speaker recognition tasks, despite the results are very close with each other. The \textit{Speech\&Phoneme-Level} approach performs best on the \textit{Phoneme-L}, achieving 71.8\% accuracy on both datasets. The \textit{Speech-Level} approach performed best on the \textit{Phoneme-1H}, achieving 80.3\% and 80.7\% accuracy on the \textit{train-clean-100} and \textit{train-clean-360} datasets, respectively. The \textit{Phoneme-Level} approach performs best on speaker recognition tasks, where it achieves 99.7\% accuracy on both \textit{Speaker-F} and \textit{Speaker-U} when using the \textit{train-clean-100} dataset. When using the \textit{train-clean-360} dataset, the \textit{Phoneme-Level} approach also achieves 99.5\% accuracy on the \textit{Speaker-U}, slightly higher than other approaches, and 99.3\% accuracy on the \textit{Speaker-F} which is on par with the \textit{Speech\&Phoneme-Level} approach.

%We illustrated the accuracy results of phoneme classification and speaker recognition with different combinations of masking approaches on Librispeech in Table~\ref{tab1}. For \textit{Mockingjay}, our proposed masking approaches have higher accuracy than random masking approach, both in phoneme classification and speaker classification tasks. Especially in phoneme classification task, the combination of speech-level and phoneme-level masking approach achieves highest accuracy on both datasets. For \textit{Tera}, our proposed masking approaches also outperform random masking approach for both tasks on \textit{train-clean-100} and for phoneme classification task on \textit{train-clean-360}. The results of proposed masking approach for speaker recognition on larger dataset is also competitive to random masking approach.

We also explored different values of proportion ratio $\rho$. The results of phoneme classification task are shown in Table~\ref{tab2}. We made quick tests ($20k$ pre-training steps, $5k$ downstream steps) on \textit{train-clean-100} dataset. Two masking approaches (\textit{Speech-Level} and \textit{Speech\&Phoneme-Level}) were investigated. The results indicated that $\rho=0.9$ is the best choice. It means $90\%$ of the masked segments are speech, and $10\%$ are non-speech. When we set $\rho$ to 0.9, both approaches showed better results than when set to several other values, in which the \textit{Speech-Level} approach achieved the accuracy of \textit{Phoneme-L} and \textit{Phoneme-1H} the highest, at 61.0\% and 68.0\%, respectively. Similarly, when the parameter is set to 0.9, the \textit{Speech-Level} approach also shows the highest accuracy of 68.0\% on the \textit{Phoneme-1H}. It also proves that some silence segments may contain high-level semantic knowledge, and they should not be discarded at all in the pre-training.

%Ablation studies over the influence of different proportion ratio parameter $\rho$ in speech-level masking approach and speech-level combined with phoneme-level masking approach were also investigated. We used \textit{train-clean-100} to pre-train our model for 20$k$ steps and the downstream task, phoneme classification, was trained by 5$k$ steps. The comparative results are shown in Table~\ref{tab2}. Apparently, $0.9$ is the best choice for proportion ratio $\rho$. This proves that some non-speech segment may also contain high-level semantic knowledge and it is not optimal to discard them at all.

\begin{table}
	\small
	\centering
	\caption{\textit{Quick Tests with Different Proportion Ratios $\rho$, Results on Librispeech, Accuracy(\%)}} 
	\vspace{1.0em}
	\setlength\tabcolsep{3pt} 
	\scalebox{0.95}{
		\begin{tabular}{|c|m{1.8cm}<{\centering}m{1.8cm}<{\centering}|m{1.8cm}<{\centering}m{1.8cm}<{\centering}|}
			
			\hline
			\multirow{2}{*}{$\rho$} & \multicolumn{2}{c|}{Speech-Level} & \multicolumn{2}{c|}{Speech\&Phoneme-Level}\\
			& {Phoneme-L} & {Phoneme-1H} & {Phoneme-L} & {Phoneme-1H} \\
			\hline
			\hline
			
			0.80 & 59.3 & 65.8 & 61.2 & 66.4\\
			0.85 & 60.2 & 67.0 & 62.1 & 66.6\\
			\textbf{0.90} & \textbf{61.0} & \textbf{68.0} & \textbf{62.3} & \textbf{68.0}\\
			0.95 & 60.0 & 65.9 & 62.2 & 67.2\\
			1.00 & 59.6 & 60.0 & 65.9 & 66.2\\
			\hline
			
	\end{tabular}}
	\label{tab2}
\end{table}

\subsection{Spectrogram Analysis}
\label{sec:analysis}

\begin{figure}[ht]
	\centering
	\subfigure[Masking Parts \newline (Random Approach)] { \label{fig2a}
		\includegraphics[width=0.45\columnwidth]{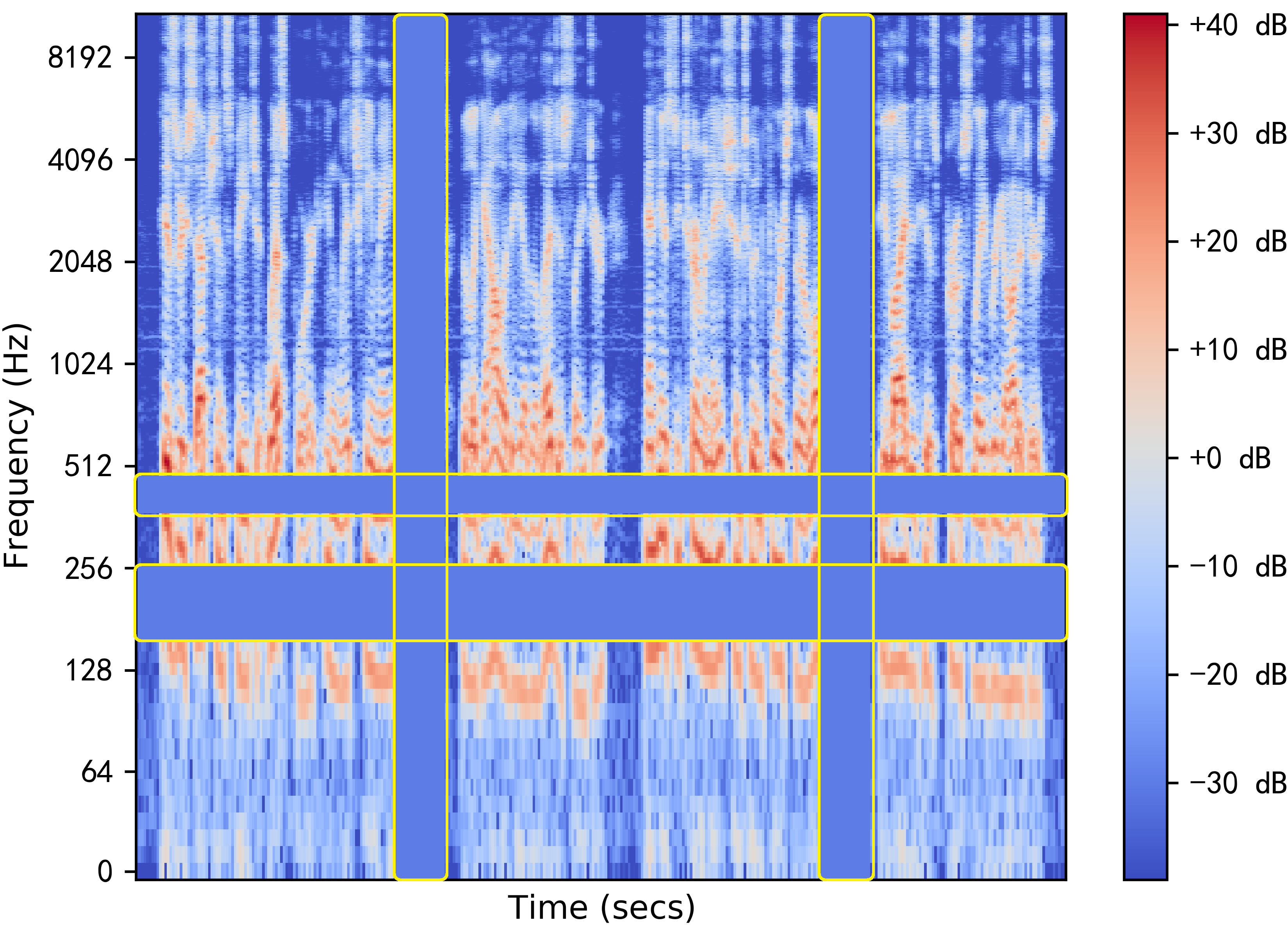}
	}
	\subfigure[Reconstructed Spectrogram \newline (Random Approach)] { \label{fig2b}
		\includegraphics[width=0.45\columnwidth]{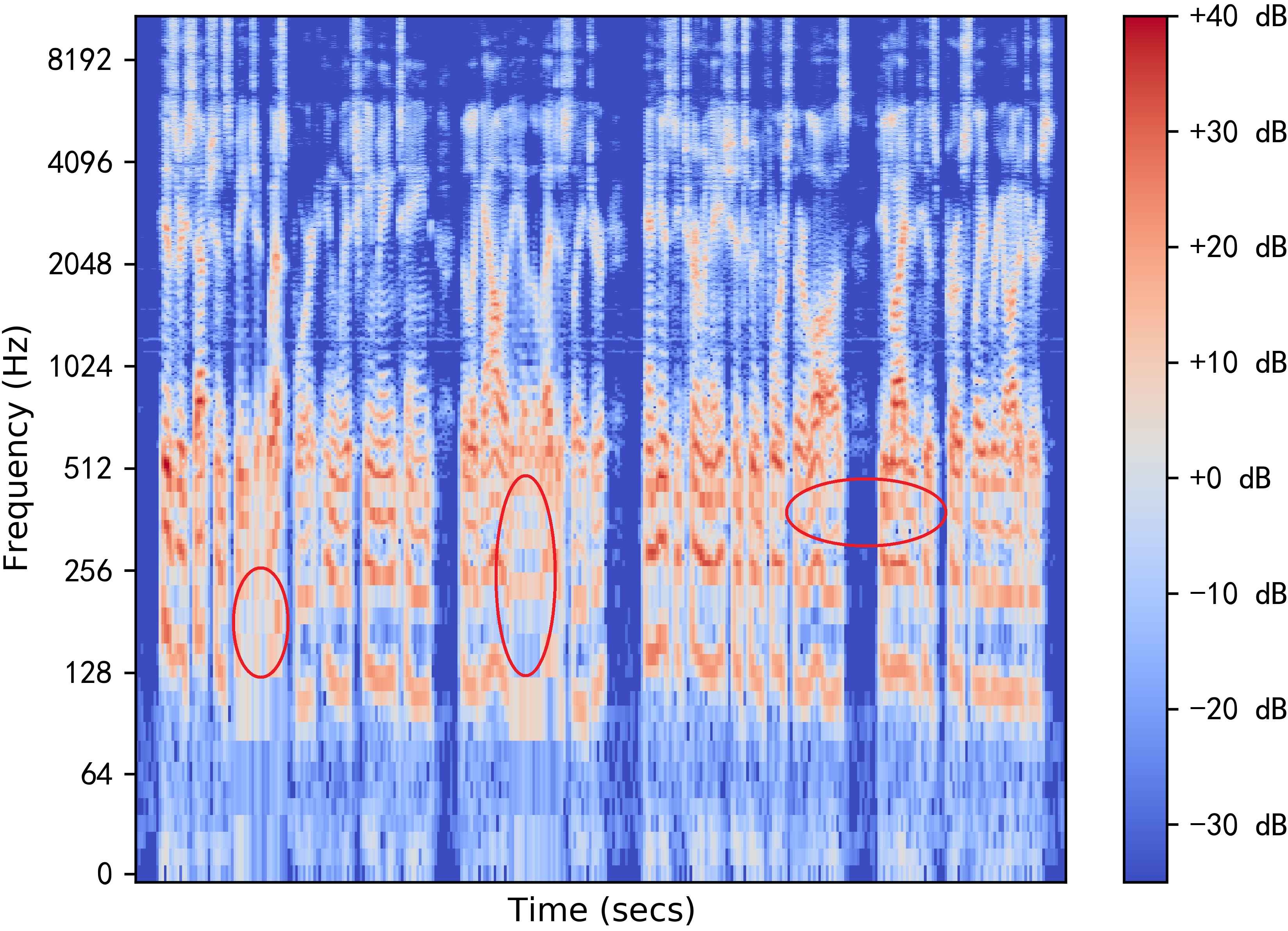}
	}
	\subfigure[Masking Parts \newline (Speech\&Phoneme-Level)] { \label{fig2c}
		\includegraphics[width=0.45\columnwidth]{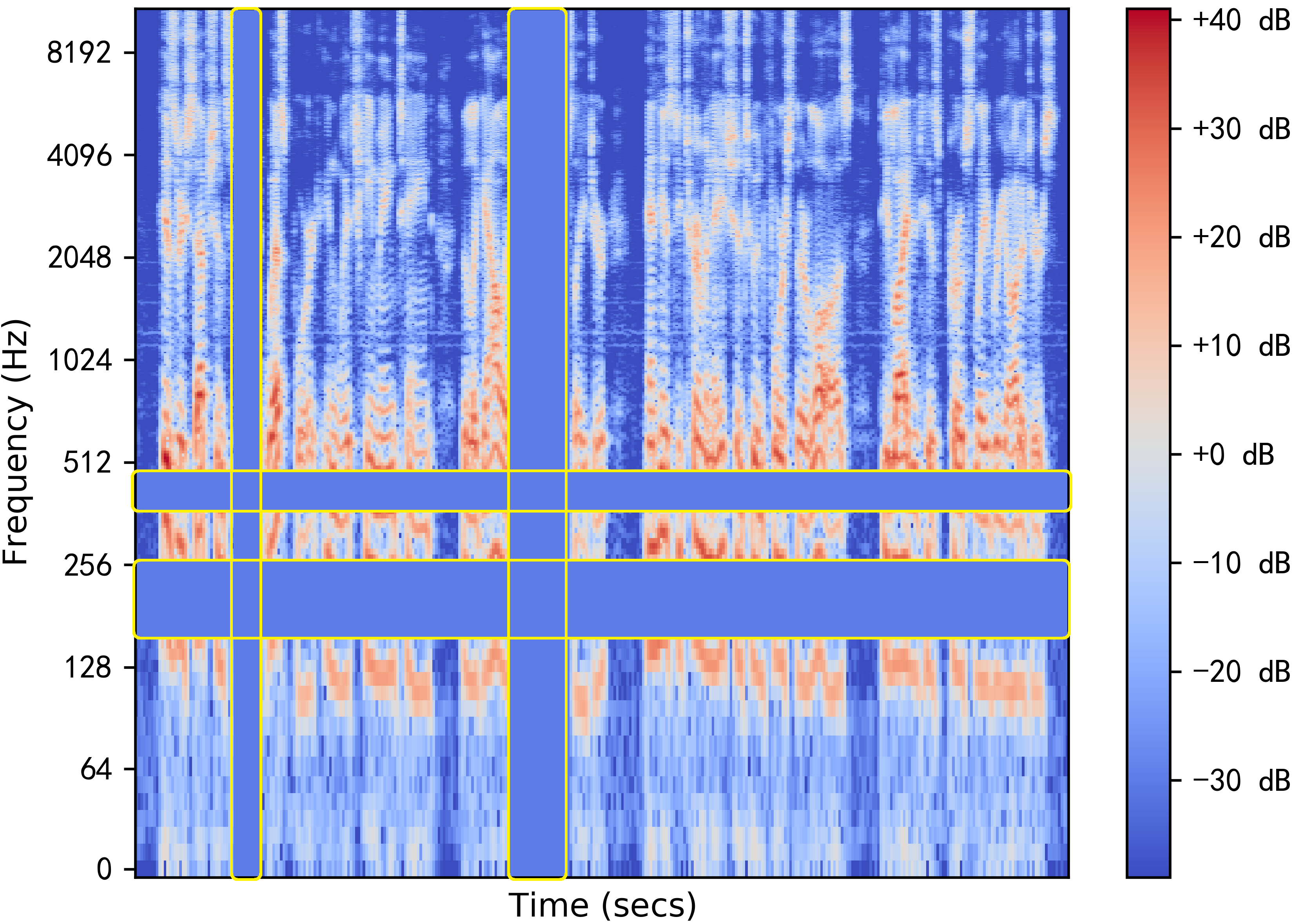}
	}
	\subfigure[Reconstructed Spectrogram \newline (Speech\&Phoneme-Level)] { \label{fig2d}
		\includegraphics[width=0.45\columnwidth]{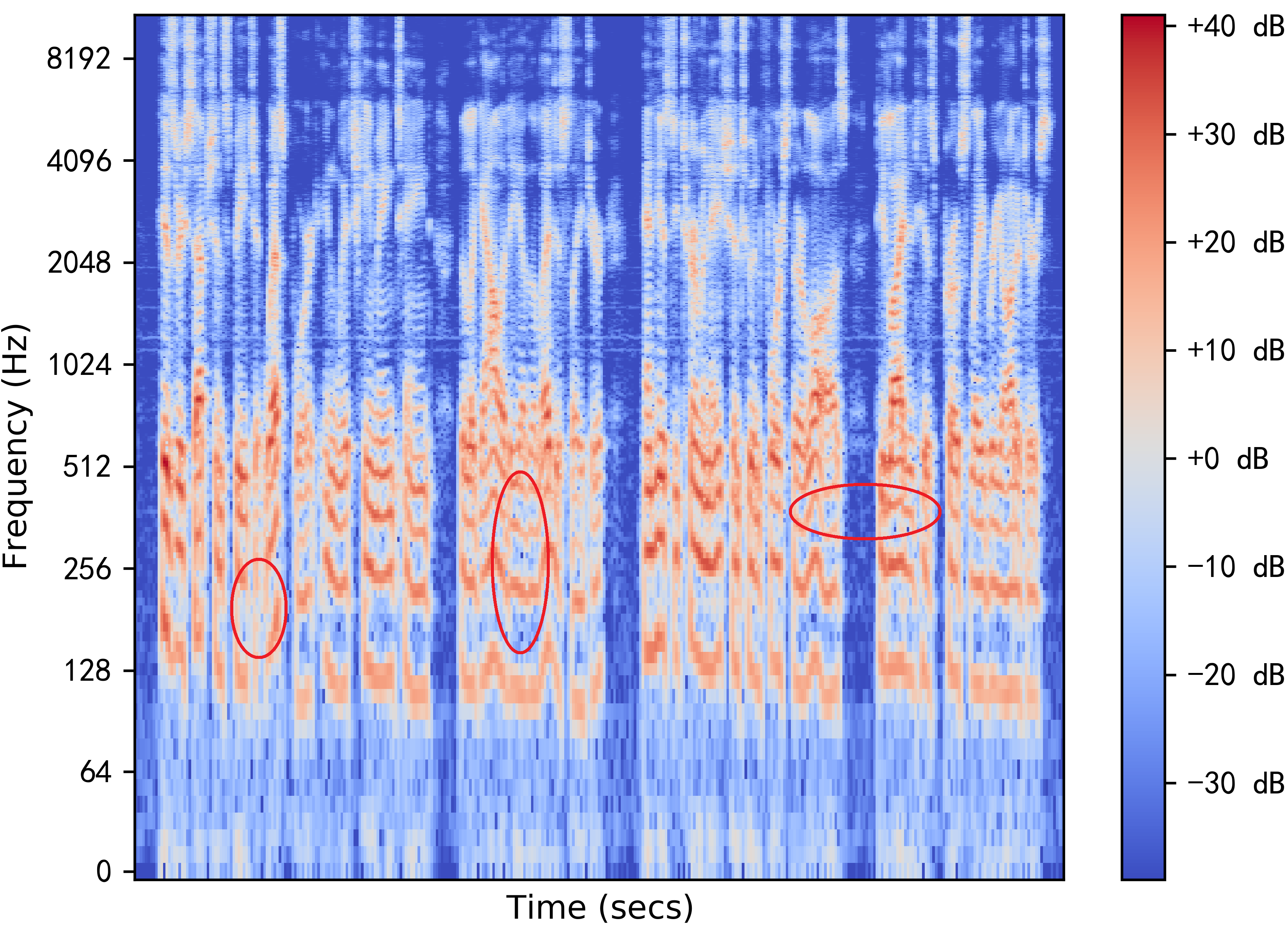}
	}
	\caption{Spectrogram Comparison Between Random and Speech\&Phoneme-Level Masking Approach.}
	\label{fig2}
\end{figure}

In this section, we plotted the masking parts of spectrogram, and reconstructed spectrogram after pre-training by TERA model. As depicted in Figure~\ref{fig2}, we made a comparison between random and \textit{Speech\&Phoneme-Level} masking approach, which are operated on one audio sample. The masking parts are highlighted in yellow lines.

For random masking, a lot of silence frames are likely masked. In addition, the masking areas have the same width in temporal dimension because the random approach masks fixed-length $C$ successive frames (in Figure~\ref{fig2a}). While for \textit{Speech\&Phoneme-Level} masking approach, the masking widths are variable, which are determined by the time duration of selected masking phonemes (in Figure~\ref{fig2c}).

After pre-training, the spectrogram is predicted, and the masking parts are supplemented. We found that the reconstructed spectrogram is over smooth for random approach (in Figure~\ref{fig2b}). It might be attributed to the local smoothness problem, which averages the surrounding signals when reconstructing the masking frames. On the contrary, our \textit{Speech\&Phoneme-Level} approach leads to a more sharp spectrogram in the masking areas (in Figure~\ref{fig2d}). It proves that the proposed methods could alleviate the over smoothness problem, and thus extract more meaningful speech representation than random approach.

\section{Conclusions}
Random masking is widely used in existing speech representation learning models. However, previous random masking method masks non-speech segments from which useful acoustic information is difficult to obtain. This work proposed two well-designed strategies, \textit{speech-level} and \textit{phoneme-level} masking approaches. The experiments show that the proposed approaches have better results on downstream tasks, than random masking. We also found that combining two masking approaches could further improve the performance. In addition, some non-speech masked segments should be reserved to provide high-level information. We could set different ratios to control the proportion of silence and speech segments. Spectrogram analysis indicated that the proposed methods could alleviate the over smoothness problem, resulting in a more sharp reconstructed spectrogram. In future works, we will investigate unsupervised method of obtaining the phoneme boundaries, instead of force-alignment, such as gate activation signal method or phoneme clustering algorithm.

\section{Acknowledgement}
This paper is supported by the Key Research and Development Program of Guangdong Province under grant No.2021B0101400003. Corresponding author is Jianzong Wang from Ping An Technology (Shenzhen) Co., Ltd (jzwang@188.com).

\bibliographystyle{IEEEtran}
\bibliography{mybib.bib}

\end{document}